\documentclass[10pt,conference]{IEEEtran}

\usepackage[dvips]{color}
\usepackage{tabu}
\usepackage{comment}
\usepackage{todonotes}
\usepackage{epsf}
\usepackage{epsfig}
\usepackage{times}
\usepackage{epsfig}
\usepackage{graphicx}
\usepackage{bbold}
\usepackage{mathtools}
\usepackage{mathrsfs}
\usepackage{amssymb,amsmath}
\usepackage{pdfpages}
\usepackage{epstopdf}
\usepackage{algorithm2e}

\usepackage{dsfont}
\usepackage{lettrine} 
\usepackage{amsmath,epsfig,amssymb,amsthm,cite,url}

\usepackage[justification=centering]{caption}
\usepackage{tabu}
\usepackage{subcaption}
\usepackage{bbm}
\def\tablename{Table}
\usepackage{algpseudocode}% http://ctan.org/pkg/algorithmicx
\algnewcommand{\LineComment}[1]{\State \(\triangleright\) #1}

\usepackage{csquotes}
\usepackage{amsmath,amssymb}
\DeclareFontFamily{OMX}{yhex}{}
\DeclareFontShape{OMX}{yhex}{m}{n}{<->yhcmex10}{}
\DeclareSymbolFont{yhlargesymbols}{OMX}{yhex}{m}{n}
\DeclareMathAccent{\wideparen}{\mathord}{yhlargesymbols}{"F3}
\usepackage{verbatim}
\usepackage{amsmath,amssymb}

\usepackage{verbatim}

\newtheorem{theorem}{\bf Theorem}
\let\oldtheorem\theorem
\renewcommand{\theorem}{\oldtheorem\normalfont}

\newtheorem{proposition}{\bf Proposition}
\let\oldproposition\proposition
\renewcommand{\proposition}{\oldproposition\normalfont}
\newtheorem{lemma}{\bf Lemma}
\let\oldlemma\lemma
\renewcommand{\lemma}{\oldlemma\normalfont}
\newtheorem{example}{Example}
\let\oldexample\example
\renewcommand{\example}{\oldexample\normalfont}

\newtheorem{definition}{\bf Definition}
\let\olddefinition\definition
\renewcommand{\definition}{\olddefinition\normalfont}
\newtheorem{remark}{Remark}
\let\oldremark\remark
\renewcommand{\remark}{\oldremark\normalfont}

\DeclareMathOperator*{\argmin}{argmin}
\usepackage{setspace}	% Remove in double column version. Also search for \setstretch in the body of the paper and comment these commands for double column
\allowdisplaybreaks

\IEEEoverridecommandlockouts
\begin{document}
% paper title
\title{\huge Optimized Path Planning for Inspection by Unmanned Aerial Vehicles Swarm with Energy Constraints\vspace*{-1em}}
\author{
	\authorblockN{Momena Monwar$^1$, Omid Semiari$^1$, and Walid Saad$^2$}\\\vspace*{-1em}
	\authorblockA{\small $^{1}$Department of Electrical and Computer Engineering, Georgia Southern University, Statesboro, GA, USA,\\ 
		Email: \protect\url{{mm16358,osemiari}@georgiasouthern.edu}\\
		$^2$Wireless@VT, Bradley Department of Electrical and Computer Engineering, Virginia Tech, Blacksburg, VA, USA, Email: \protect\url{walids@vt.edu}
	}\vspace*{-2em}
	 \thanks{This research was supported by the  U.S. National Science Foundation under Grants OAC-1541105 and OAC-1638283.
	}%
}
%
% make the title area
%\vspace{-1em}
%
\maketitle
\begin{abstract}
Autonomous inspection of large geographical areas is a central requirement for efficient hazard detection and disaster management in future cyber-physical systems such as smart cities. In this regard, exploiting unmanned aerial vehicle (UAV) swarms is a promising solution to inspect vast areas efficiently and with low cost. In fact, UAVs can easily fly and  reach inspection points, record surveillance data, and send this information to a wireless base station (BS). Nonetheless, in many cases, such as operations at remote areas, the UAVs cannot be guided directly by the BS in real-time to find their path. Moreover, another key challenge of inspection by UAVs is the limited battery capacity. Thus, realizing the vision of autonomous inspection via UAVs requires \emph{energy-efficient path planning} that takes into account the energy constraint of each individual UAV. In this paper, a novel path planning algorithm is proposed for performing energy-efficient inspection, under stringent energy availability constraints for each UAV. The developed framework takes into account all aspects of energy consumption for a UAV swarm during the inspection operations, including energy required for flying, hovering, and data transmission. It is shown that the proposed algorithm can address the path  planning problem efficiently in polynomial time. Simulation results show that the proposed algorithm can yield substantial performance gains in terms of minimizing the overall inspection time and energy. Moreover, the results provide guidelines to determine parameters such as the number of required UAVs and  amount of energy, while designing an autonomous inspection system.
 \vspace{-0cm}
\end{abstract}
\section{Introduction}\label{intro} \vspace{-0cm}
Autonomous inspection of large geographical areas is a key requirement to increase safety and optimize the operation of future cyber-physical systems, such as smart cities, smart farms \cite{WOLFERT201769}, and smart oil fields \cite{sof2}. In smart farms, for example, there is a need to inspect large farm lands for signs of pest attacks and to spray pesticide locally over vulnerable regions.  Meanwhile, smart oil fields require constant monitoring of the fields for detecting signs of oil spills. Another prominent application of autonomous inspection is in post-disaster management, after natural calamities like hurricanes or floods, to perform search-and-rescue operations.

In this regard, inspection by \emph{unmanned aerial vehicles (UAVs)} \cite{7412759} is seen as an attractive solution due to the following reasons: 1)  UAVs can fly and bypass obstacles which can result in reducing the inspection time; 2) UAVs are relatively cheaper than dispatching human personnel; and 3) UAVs can easily operate in adverse location that can be dangerous to human workers. For instance, a byproduct of oil extraction is Hydrogen Sulphide (H2S), a colorless and odorless toxic gas that makes oil spill inspection challenging.
In such scenarios, a group of UAVs, also known as a \emph{UAV swarm}, can perform coordinated inspection and manage the inspection tasks more efficiently with a low cost and within a limited time. 
%UAV swarms  Each UAV will have certain path to travel and monitor predetermined points in order to gathering information or delivering task or providing wireless connectivity to the landscape of wide-area where connectivity was sparing or has been compromised due to extreme geographical condition.    

%Recent years have witnessed, a significant amount of research carried on the usage of interconnected multiple UAVs in an adaptive and autonomous acting system. A UAV swarm is capable of accomplishing tasks that is much more challenging for one UAV to fulfill. For example, accurate determination of the location for an object, mapping of inaccessible Ocean, assessment of real-time environmental processes, or agricultural and industrial monitoring. Furthermore, a UAV swarm is able not only to solve tasks, but also to reduce the accomplish time providing a better quality of collected data. Additionally, if the task requires navigational autonomy within an unknown or difficult environment, a UAV swarm offers robustness through redundancy and self-organization, which cannot be achieved by deploying one UAV \cite{Danoy:2015:CSA:2815347.2815351}.

%UAV swarms can be leveraged for post-disaster management, after natural calamities like hurricanes or floods to explore large areas for finding victims. Each UAV will have certain path to travel and monitor predetermined points in order to gathering information or delivering task or providing wireless connectivity to the landscape of wide-area where connectivity was sparing or has been compromised due to extreme geographical condition. 

Nonetheless, several challenges must be addressed to realize practical autonomous inspection by a UAV swarm. \emph{First}, each UAV has a limited battery capacity that can support a limited flight time. Thus, UAVs must work collaboratively to inspect a large geographical area. \emph{Second}, a UAV swarm may include heterogeneous types of UAVs with different flying capabilities and energy efficiency. \emph{Third}, in many scenarios, UAVs cannot be fully controlled by a wireless base station (BS) to find their path, e.g., in post-disaster situations where the communication infrastructure is damaged. \emph{Fourth}, as the UAV's distance from the BS increases, more transmission power is required by the UAV to maintain a minimum data transmission rate. \emph{Finally}, in most applications, the inspection must be performed in a limited time. For example, in oil spill detection, a few minutes saving in the inspection time can prevent drastic oil spill incidents. Therefore, reaping the full benefits of UAV swarms for autonomous inspection mandates efficient \emph{path planning} for each UAV to minimize the overall energy consumption, while accounting for the limited energy of each UAV.

Recently, several works have been proposed in the literature to address path planning for autonomous inspection by a UAV swarm \cite{PP1,PP2,7888557,7192644,8255824, PP4,8243572}. The authors in \cite{PP1} propose an age-optimal trajectory planning for UAVs  to collect data from ground sensor nodes. In \cite{PP2}, a framework is developed to exploit UAVs, serving as relay nodes, to minimize the peak age-of-information for a pair of nodes in an Internet-of-Things (IoT) network. The authors in \cite{7888557} find an energy-efficient path planning scheme for a fixed-wing UAV that communicates with a ground user. Moreover, the work in \cite{7192644} studies an energy-efficient cooperative relaying by a set of UAVs, while considering circular flight trajectories. Meanwhile, \cite{8255824} proposes a scheme to jointly optimizing the UAV's trajectory and transmit power allocation.  The work in \cite{PP4} presents a tutorial on the applications of UAVs in wireless communications  and studies fundamental tradeoffs in UAV-enabled wireless networks. With regard to optimizing UAV deployments for inspection applications, \cite{8243572} presents a path planning scheme for the inspection of photovoltaic farms by a UAV swarm.

%In \cite{7106967}, the problem of establishing an efficient swarm movement model and a network topology between a collection of UAVs, is solved via implementing a leader-election algorithm to a set of autonomous micro UAVs. In this local decision making approach, the leader will collect information from the swarm and lead the swarm to the destination, while maintaining communication with the base station \cite{7106967}. 
%In \cite{7888557}, the authors derive an energy consumption model from real measurements to optimize the UAVs' coverage area and maximize communications energy efficiency. 
% In \cite{DBLP:journals/corr/abs-1802-06042}, a self-organizing LTE-based communication network is proposed for managing low-altitude UAVs. 
% Moreover, in \cite{ref2}, the authors propose a solution for coalition formation problem in a heterogeneous resource-constraint UAV network to maximize the reliability by selecting the most trustworthy UAVs among the available self-interested UAVs in the network. Moreover, several works focus on UAV applications in wireless communication, mainly as a temporary BS, for coverage extension and capacity enhancement \cite{12-13}.

%Cooperation formation has been a very active area of research in multi-agent systems. Many researchers have attempted to deal with the problem of efficient network formulation with multi-repairmen systems by applying various approaches including dynamic programing methods \cite{z}, graph theory \cite{Bistaffa:2014:ACS:2615731.2615737,Sless:2014:FCF:2615731.2615776}, iterative processes \cite{7817072}. 

Although interesting, the body of work in \cite{PP1,PP2,7192644,8255824,PP4} mainly focuses on communications aspects of UAVs, without considering specific constraints that affect inspection operations, such as the impact of UAV's weight, energy requirements for hovering, and flight conditions. Moreover, some of the existing works consider path planning for specific scenarios, such as the works in \cite{7888557} and \cite{7192644}, that solely focus on circular trajectory for the UAVs. In addition, \cite{8243572} does not consider communications constraints such as minimum required data rate. Meanwhile, several works in the literature, such as \cite{PP1,PP2,7192644} and \cite{8255824}, primarily focus on minimizing the time that a wireless user is visited by the UAV swarm. While this metric is important in sensor networks and IoT, for inspection operations, energy-efficient path planning is more critical.  Additionally, missing from the prior art, there is a need for performance analysis to determine parameters such as the number of required UAVs and energy requirements for inspection operations by a UAV swarm.

The main contribution of this work is a novel path planning algorithm to minimize the overall energy consumption for inspection by a UAV swarm, while accounting for the heterogeneous features of UAVs and their unique energy constraints. In particular, we first propose an energy consumption model that jointly captures the required energy for flying, hovering, and transmission of recorded inspection data by a UAV to the wireless BS. To solve the autonomous inspection problem, we model the area as an undirected, weighted graph with critical inspection points as the vertices of the graph and the required energy for monitoring an inspection point as the weight of the edge. Using the graph-based model, we develop a novel path planning algorithm that exploits the analogies between  autonomous inspection and the $K$-traveling repairmen problem \cite{PP6}, while considering the limited energy budget of each UAV. In particular, the proposed algorithm jointly optimizes the trajectory of all energy-constrained UAVs in a swarm, in order to minimize the overall average inspection energy. Simulation results show that the proposed algorithm outperforms the distance-based path planning, both in terms of required time and energy for inspection of an area. The results also shed light on design and performance analysis of the autonomous inspection by a UAV swarm, in terms of the number of required UAVs, average inspection time, and energy consumption.

The rest of paper is organized as follows. Section II presents the system model and problem formulation. Section III presents the proposed solution. Simulation results are provided in Section IV. Section V concludes the paper.

\begin{figure}[t!]
	\centering
	\centerline{\includegraphics[width=5.5cm]{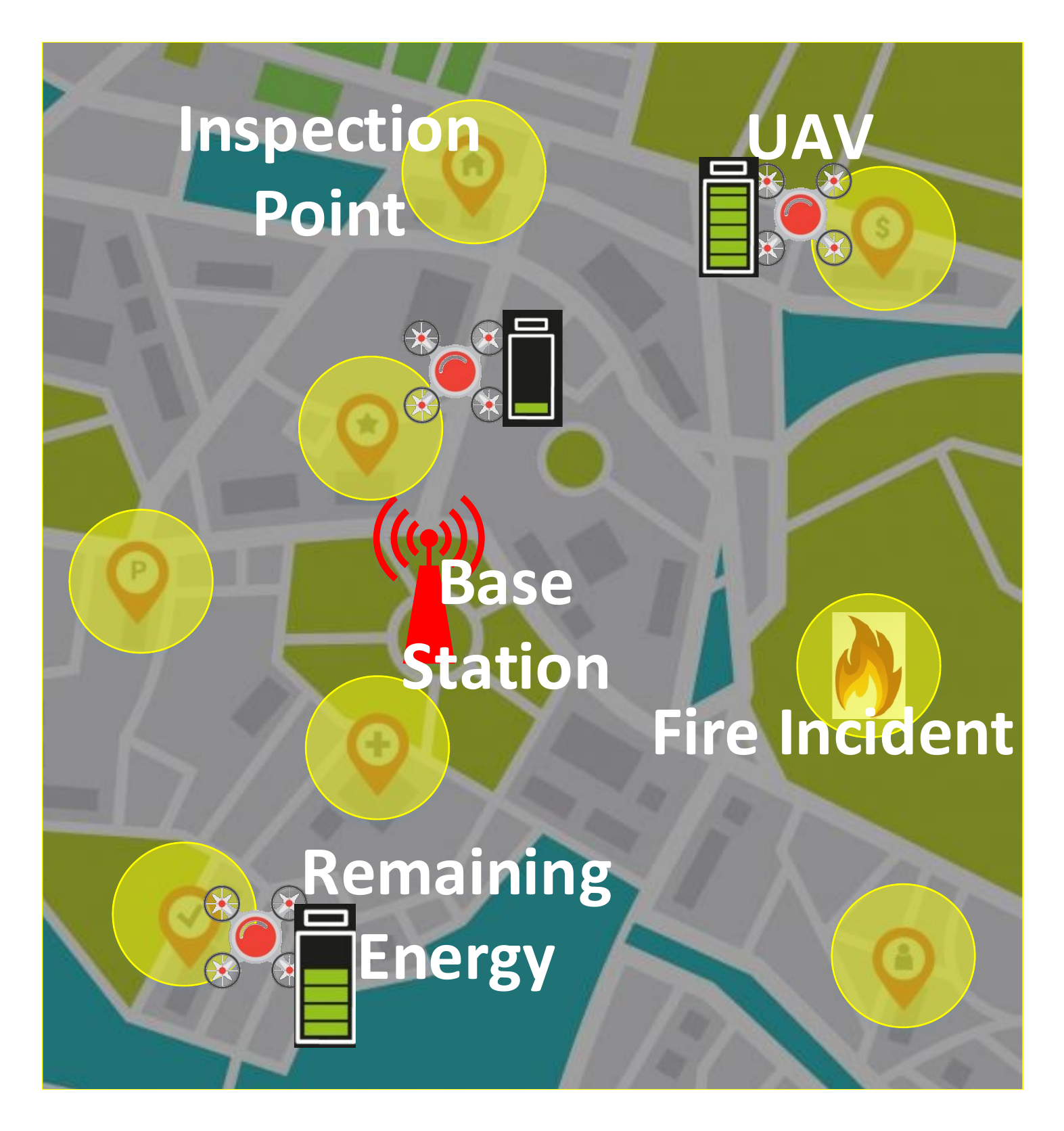}}\vspace{-1em}
	\caption{\small Example scenario for inspection with a UAV swarm.}\vspace{-.2cm}
	\label{model}
\end{figure} 
\section{System Model}
Consider a swarm of $K$ heterogeneous UAVs -- with different energy efficiency and battery size -- in a set $\mathcal{K}$ that are used to perform autonomous inspection of a square area. Fig. \ref{model} shows an example scenario of autonomous inspection by a UAV swarm. A wireless BS, located at the origin $\boldsymbol{x}_o=(0,0)\in \mathbbm{R}^2$, also serves as the UAVs' base where the UAVs start their travel. Each UAV $k \in \mathcal{K}$ is constrained with an energy budget $E_k$. In addition, let $\mathcal{N}$ be the set of $N$ inspection points that must be visited by at least one UAV during a time window, determined by the actual inspection application. For example, for post-disaster rescue operations, the inspection time may not exceed few minutes. Next, we find the energy consumption of each UAV as a function of different parameters, such as the location, energy efficiency, and traversed distance. Without loss of generality, we focus on quadrotor UAVs, while other types of UAVs can also be taken into account.

Energy consumption for monitoring of an inspection point by a UAV includes three components: 1) \emph{Flight energy}: the required energy to fly to the point of interest from another location; 2) \emph{Transmission energy}: the required energy to transmit the recorded data (photo or video) from the inspection point to the BS; and 3) \emph{Hovering energy}: the energy consumption for remaining still on the air during the time that takes for a UAV to record and transmit a data. Next, we characterize each energy consumption component.

\subsection{Flying and Hovering Energy Consumption}

The energy consumption by a flying UAV is mainly spent to overcome the gravity for staying in the air, as well as drag forces due to wind and forward motions. Building on the model in \cite{PP5}, the minimum power with forward motion for a UAV is given by
\begin{align}\label{eq1}
	p_{\text{min}} = (\hat{v} + v\sin{\beta})T,
\end{align} 
where $\hat{v}$ is the induced velocity required for a given thrust $T$, $v$ is the average ground speed of the UAV, and $\beta$ denotes the pitch angle. For a UAV with mass $m$, including the mass of the UAV body and battery, the required thrust $T$ is:
\begin{align}\label{eq2}
	T = mg + f_d,
\end{align} 
where $g$ is the gravitational constant and $f_d$ is the drag force that depends on the air speed, density of air $\rho$, and the drag coefficient. Given a thrust $T$, the induced velocity can be calculated by solving the following nonlinear equation \cite{PP5}:
\begin{align}\label{eq3}
	\hat{v} = \frac{2T}{q r^2\pi \rho \sqrt{(v\cos \beta)^2+(v \sin \beta + \hat{v})^2}},
\end{align} 
where $r$ and $q$ represent, respectively, the diameter and number of UAV rotors. The theoretical power consumption in \eqref{eq1} can be used to find the actual power consumption by a UAV $k \in \mathcal{K}$ as:
\begin{align}
	p_f(k) = \frac{p_f^{\text{min}}}{\eta_k},
\end{align}
where $\eta_k$ is the power efficiency of the UAV $k$, known empirically. With that in mind, the total flying energy that an arbitrary UAV $k$ consumes to traverse distance $d$ is:
\begin{align}\label{Ef}
	E_f(k) = p_f(k)\frac{d}{v} = \frac{p_f^{\text{min}}d}{v\eta_k}.
\end{align}

Furthermore, the actual power consumption for hovering of UAV $k$ is \cite{PP5}:
\begin{align}\label{hover_p}
	p_h(k)=\frac{p_h^{\text{min}}}{\eta_k}=\frac{T\sqrt{T}}{\eta_k\sqrt{0.5\pi q r^2 \rho}},
\end{align}
where $p_h^{\text{min}}$ is the theoretical minimum power for hovering. 
\subsection{Transmission Energy Consumption}
The energy consumption for transmission of recorded data from the inspection point cannot be ignored in many cases where: 1) The distance between the UAV and the BS is large; and 2) The size of the recorded content is large (such as high-definition videos). Considering the inspection of a point at $\boldsymbol{x}=(x_1,x_2) \in \mathbbm{R}^2$, the path loss of the wireless link in dB is given by:
\begin{align}\label{path loss}
	L = L_0 + \alpha \log_{10}\left(\sqrt{x_1^2 + x_2^2}\right) + \xi,
\end{align}
where $L_0$ is the path loss at a reference distance, $\alpha$ is the path loss exponent, and $\xi$ represents the shadowing effect in dB scale. Using \eqref{path loss}, we can find the achievable data rate of the uplink between a UAV $k$ and the BS as follows\footnote{Although we consider log-distance path loss channel model, small-scale fading can also be easily accommodated in \eqref{rate}.}:
\begin{align}\label{rate}
	R_k = w_k \log_2 \left(1 + \frac{p_t(k)10^{-L/10}}{w_k N_0}\right),
\end{align}
where $w_k$, $p_t(k)$, and $N_0$ denote, respectively, the bandwidth, the transmit power of UAV, and the noise power spectral density. In \eqref{rate}, orthogonal bandwidth allocation is considered across UAVs, thus, there is no interference between simultaneous UAV transmissions. To find the transmission energy consumption, let $R_{\text{th}}$ be the minimum required uplink data rate for successful transmission of the recorded content. From \eqref{rate} and subject to the data rate requirement $R_{\text{th}}$, one can easily find the minimum required transmit power $p_t^{\text{min}}(k)$ for a UAV $k$. Meanwhile, the energy consumption for transmission of a data packet with size $B$ bits will be:
\begin{align}\label{Et}
	E_t(k) = \tau\frac{p_t^{\text{min}}(k)}{\eta_k} = \frac{Bp_t^{\text{min}}(k)}{R_{\text{th}}\eta_k},
\end{align}
where $\tau$ is the over-the-air transmission latency. From \eqref{Ef}-\eqref{Et}, we can conclude that the overall energy consumption by a UAV $k \in \mathcal
K$ to fly from the base, located at $n_0$,  to inspect a destination point $n \in \mathcal{N}$ is:
\begin{align}\label{energy}
E_k^{\text{total}}(n)&= E_k^{\text{total}}(n') + E_t(k)+E_h(k)+E_f(k) \notag\\
&= E_k^{\text{total}}(n') + \frac{B}{R_{\text{th}}\eta_k} \left[p_t^{\text{min}}(k)+ p_h^{\text{min}}(k)\right] + \frac{p_f^{\text{min}}d}{v\eta_k},
\end{align}
where $n'$ shows the inspection point visited immediately before $n$. In \eqref{energy}, $d$ is the distance between points $n$ and $n'$ and $p_t^{\text{min}}(k)$ depends on the location of inspection point $n$. In fact, given that $E_k^{\text{total}}(n_0)=0$, \eqref{energy} allows to recursively find the overall inspection energy consumption for any arbitrary point.
\subsection{Problem Formulation}
To study the path planning problem, we model the inspection area as an undirected, weighted graph $G_k(\mathcal{N}, \mathcal{E}_k)$, corresponding to the UAV $k$, where the vertices are inspection points and the weight of each edge between points $(n',n)$ is $E_k^{\text{total}}(n)-E_k^{\text{total}}(n')$ and can be found from \eqref{energy}. The \emph{goal} here is to find the best trajectory for each UAV across the inspection points (i.e., path planning) in order to minimize the overall inspection time. The union of these trajectories must cover all the inspection points in the set $\mathcal{N}$. Prior to formulating the problem, we use the following definitions:

\definition{The \emph{trajectory} of a UAV $k$ is defined as a tree $\mathcal{T}_k$, inside the graph $G_k(\mathcal{N}, \mathcal{E})$, with the root starting at the location of the BS, $n_0$.} 

The main reason to define each UAV's trajectory as a tree is to avoid cycles within the inspection path. Associated with each trajectory, we can define a cost in terms of energy consumption as follows:
\definition{The \emph{cost} of a trajectory that visits inspection points $\mathcal{T} =\{ n_0, n_1, n_2, \cdots, n_j\}$, for UAV $k$ is defined as the total energy required to inspect the vertices of $\mathcal{T}$. That is, 
\begin{align}\label{cost}
	C(\mathcal{T};k) = \sum_{i=0}^{j}E_k^{\text{total}}(n_i).
\end{align}}Even though other energy cost functions can be defined here, the advantage of the cost function in \eqref{cost} is to yield a unique cost for each tree, irrespective of the order that UAV may visit vertices.  With this in mind, we aim to find $K$ trajectories $\mathcal{T}_k$, for $k=1,2,\cdots, K$, that minimize the overall inspection energy consumption. That is,
\begin{align}
\argmin_{\mathcal{T}_k, k=1,2,\cdots, K} \,\,\,&\sum_{k \in \mathcal{K}} C(\mathcal{T}_k;k),\label{opt1}\\
\text{s.t.,}& \bigcup\limits_{k=1}^{K} \mathcal{T}_{k} = \mathcal{N}, \label{opt2}\\
&\bigcap\limits_{k=1}^{K} \mathcal{T}_{k} = \{n_0\},\label{opt3}\\
&C(\mathcal{T}_k;k) \leq E_{\text{th}}(k),\label{opt4}
\end{align}
where constraint \eqref{opt2} ensures that all inspection points are visited. Particularly, \eqref{opt2} implies that the trajectories of UAVs are interdependent, thus, path planning must be done jointly for all UAVs in the swarm. Constraint \eqref{opt3} indicates that trajectories of UAVs do not overlap, except at the starting point $n_0$. In addition, \eqref{opt4} ensures that the total energy consumption by a UAV $k$ does not exceed the energy budget $E_{\text{th}}(k)$. Prior to solving this problem, we make the following observation with regard to the complexity of the problem:

\proposition{The proposed path planning problem in \eqref{opt1}-\eqref{opt4} is NP-hard and cannot be solved optimally within polynomial time.} 

\begin{proof}
To show this, we note that if constraint \eqref{opt4} is relaxed, i.e., $E_{\text{th}}(k) \rightarrow \infty$, for all $k \in \mathcal{K}$, then the proposed problem will be analogous to the $K$-traveling repairmen problem which is known to be NP-hard \cite{PP6}. Thus, with finite energy budget, the proposed problem is \emph{reducible} to the $K$-traveling repairmen problem, and thus, it is also NP-hard.
\end{proof}

Here, we note that classic solutions, such as Dijkstra's method, for solving the traveling salesman problem cannot be used to solve \eqref{opt1}-\eqref{opt4}, as these approaches consider only one agent (UAV) and in addition, cannot capture constrains in \eqref{opt4}. With this in mind, next, we will propose an efficient path planning algorithm that aims to minimize the overall energy consumption, while taking into account the energy constraint for each UAV.

\section{Proposed Path Planning Algorithm}
Prior to developing an efficient approximation algorithm for the proposed problem in \eqref{opt1}-\eqref{opt4}, lets consider the following simple scenario which we will use as a subroutine for our algorithm.

Assume that there is only one UAV $k'$. Given the graph $G_{k'}(\mathcal{N}, \mathcal{E}_{k'})$, and starting the trajectory from  $n_0$, we aim to find the least expensive tree $\mathcal{T}_{k'}$ (cost is calculated from \eqref{cost})  that has exactly $j$ vertices. This problem is known as \emph{j-minimum spanning tree ($j$-MST)} which is also NP-hard \cite{PP6}. Although approximation algorithms are available to find $j$-MST efficiently, we need to modify this problem to accommodate the limited energy budget of the UAV (as presented in \eqref{opt4}). In this regard, we consider the subroutine algorithm presented in Table. \ref{algo1}, which is originally proposed in \cite{PP6} and allows to convert the cardinality constraint of covering exactly $j$ vertices into an energy budget constraint.  From the definition of MST, we can immediately conclude the following:

\begin{table}[t!]
	\small
	\centering
\caption{\small Approximation Algorithm for the $j$-MST Problem with Energy Constraint}\vspace*{-0.2cm}
	\begin{tabular}{p{8 cm}}
		\hline \vspace*{-0em}
		\textbf{Inputs:}\,\,$G_{k'}(\mathcal{N}, \mathcal{E}_{k'}), E_{\text{th}}(k')$.\\
		\hspace*{1em}
		\textbf{Initialize:} Let $\mathcal{T}_{k'}=\emptyset$.\\
		%%%%%%%%%%%%%%%%%%%%%%%%%%%%%%%%%%%
		\hspace*{-0.5em} \For{$i=1$ to $j$}{
			\textbf{Step1:} Find the MST for $i$ vertices, using Prim algorithm. This yield an approximation for the $i$-MST problem. Denote the MST as $Y(i)$. 
			
				\textbf{Step2:} For a constant $\lambda \geq 1$,
			
			\If{$C(Y(i);k') \leq \lambda E_{\text{th}}(k')$}{Let $\mathcal{T}_{k'}=Y(i)$.}
} 				
		\hspace*{0em}\textbf{Output:}\,\, $\mathcal{T}_{k'}(E_{\text{th}})$\vspace*{0em}\\
		\hline
	\end{tabular}\label{algo1}\vspace{-1em}
\end{table}
\begin{remark}
If there exists a tree of cost $E_{\text{th}}(k)$ that spans $i$ vertices, the algorithm in Table \ref{algo1} always returns a tree with a cost less than $\lambda E_{\text{th}}(k)$ that covers $i$ vertices.
\end{remark}
%\begin{proof}
%	The proof directly results from the definition of MST.
%\end{proof}

Using the algorithm adopted in Table \ref{algo1}, we propose a new algorithm in Table \ref{algo2} that yields an efficient solution for the path planning problem of a UAV swarm with energy constraints. The proposed algorithm proceeds as follows. Initially, the algorithm sorts all UAVs with respect to their available energy $E_{\text{th}}$ in an ascending order. Starting from the minimum energy $E_0$, at each round, the algorithm aims to cover all inspection points in $\mathcal{N}$ by finding $K$ trees. If such set of trees do no exist for the given energy constraint, in the next round, the energy is increased by $\Delta E$. Given a certain energy budget $E_i$ at round $i$, in step 2, the algorithm uses the subroutine in Table \ref{algo1} to find $\mathcal{T}_k(E_i)$. In step 3, the algorithm tentatively assigns $\mathcal{T}_k(E_i)$ as the path for the UAV $k$. If the further increase in the energy budget $E_i$ exceeds the available energy of the UAV $E_{\text{th}}(k)$, the algorithm returns $\mathcal{T}_k$ in step 4 and removes the UAV $k$ and the subset of assigned vertices from the graph. Otherwise, in step 5, the vertices of $\mathcal{T}_k(E_i)$ are tentatively removed from $\tilde{\mathcal{N}}$.
\proposition{The proposed algorithm in Table \ref{algo2} converges in polynomial time.}
\begin{proof}
	Lets consider $E_{\text{th}}(k)=\infty$, for all UAVs. Clearly, this case provides an upper bound for the number of iterations in our algorithm, since with infinite energy, the condition in step 4 is never satisfied and no UAV is permanently removed from the algorithm. In this case, given that at each round of the algorithm in Table \ref{algo2} the energy budget increases (equivalent to increasing $j$ in the $j$-MST), the proposed algorithm with converge after maximum $N$ iterations. Moreover, within each round of the proposed algorithm, there are $K$ recalls of the subroutine algorithm in Table \ref{algo1}. Meanwhile, we note that this algorithm solves the $j$-MST problem and runs in polynomial time.  Therefore, the algorithm is guaranteed to converge in polynomial time.
\end{proof}

\begin{table}[t!]
	\small
	\centering
	\caption{\small Proposed Path Planning Algorithm for the UAV Swarm}\vspace*{-0.2cm}
	\begin{tabular}{p{8 cm}}
		\hline \vspace*{-0em}
		\textbf{Inputs:}\,\,$G_{k}(\mathcal{N}, \mathcal{E}_k), E_{\text{th}}(k)$, for all $k=1,2,\cdots,K$, $\Delta E$.\\
		\hspace*{1em}
		\textbf{Initialize:} Let $\mathcal{T}_{k}=\emptyset$, for all $k=1,2,\cdots,K$, $E_0 = \Delta E$. Sort the UAVs, based on $E_{\text{th}}(k)$, in ascending order in a list $\tilde{\mathcal{K}}$.  Let $\hat{\mathcal{N}}=\emptyset$.\\
		%%%%%%%%%%%%%%%%%%%%%%%%%%%%%%%%%%%
		\hspace*{-0.5em} \While{$\tilde{\mathcal{N}}\neq\{n_0\}$}{
			$\tilde{\mathcal{N}} \leftarrow \mathcal{N}$
			
			\textbf{Step1:}  Remove vertices in $\hat{\mathcal{N}}$ from $\tilde{\mathcal{N}} $.
			
			\For{$k'=1$ to $|\tilde{\mathcal{K}}|$}{
			\textbf{Step2:} Let $k$ be the index of $k'$-th UAV from the sorted list $\tilde{\mathcal{K}}$. Using the algorithm in Table. \ref{algo1}, find the tree $\mathcal{T}_{k}(E_i)$ in $G_{k}(\tilde{\mathcal{N}}, \mathcal{E})$.

			\textbf{Step3:} \If{$C(\mathcal{T}_{k}(E_i);k) \leq  E_{\text{th}}(k)$}{Let $\mathcal{T}_{k}=\mathcal{T}_{k}(E_i)$.}
			
				\textbf{Step4:} 
			\If{$C(\mathcal{T}_{k}(E_i);k) + \Delta E  > E_{\textrm{th}}(k)$}{Add vertices of $\mathcal{T}_{k}(E_i)$ to the set $\hat{\mathcal{N}}$. Remove $k$ from the list $\tilde{\mathcal{K}}$. Return $\mathcal{T}_{k}$.}
			
			\textbf{Step5:} Remove the vertices of $\mathcal{T}_{k}(E_i)$, except $n_0$, from $\tilde{\mathcal{N}}$.

}	
$E_{i+1} \leftarrow E_i + \Delta E$		
		} 				
		\hspace*{0em}\textbf{Output:}\,\, $\mathcal{T}_{k}$, for $k=1,2,\cdots,K$.\vspace*{0em}\\
		\hline
	\end{tabular}\label{algo2}\vspace{-1em}
\end{table}

\vspace{-1em}
\section{Simulation Results}

We consider an inspection area of size $200 \times 200$ meters with the BS located at $\boldsymbol{x}_0=(0,0)$ and inspection points are distributed uniformly and randomly across the inspection area. Simulation parameters and their description are summarized in Table \ref{tabsim}. The statistical results are averaged over a large number of simulation runs. 

We compare the performance of our proposed approach with a heuristic path planning algorithm, called \emph{distance-based trajectory}, in which each UAV starts from a random point and at each step, moves to its nearest inspection point in the graph. This approach is analogous to the ``nearest neighbor algorithm'' for solving the traveling salesman problem. In fact, the baseline algorithm does not incorporate the energy constraints of the UAVs while performing the path planning. In addition to the inspection energy consumption, we consider the inspection time, as another important performance metric, which is defined as follows: Given paths $\mathcal{T}_k$ for all UAVs $k=1,2,\cdots,K$, the \emph{inspection time} of an area is the overall time that takes for all inspection points to be visited by at least one UAV for the first time.
\begin{table}[t!]
	\scriptsize
	\centering
	\caption{%\mycaption{%\vspace*{-1em}
		\vspace*{-0cm} Simulation Parameters }\vspace*{-0.2cm}
	\begin{tabular}{|c|c|c|}
		\hline
		\bf{Notation} & \bf{Parameter} & \bf{Value} \\
		\hline
		-- & mass of UAV & $1.07$ kg \cite{PP5}\\
		\hline
		-- &  mass of battery & $1$ kg \cite{PP5}\\
		\hline
		$\rho $ & density of air  & $1.225$ kg/m$^3$ \cite{PP5}\\
		\hline
		$f_{d}$ & drag	force & $9.6998$ N \cite{PP5}\\
		\hline
		$q$ & number	of	rotors & $4$ \cite{PP5}\\
		\hline
		$r$ & rotor diameter & $0.254$ m \cite{PP5}\\
		\hline
		$\eta$ & power efficiency & $70\%$ \cite{PP5}\\
		\hline
		$v$ & average ground speed & $1.49$ $m/s$ \cite{PP5}\\
		\hline
		$B$ & data packet size & $20$ Mbits \\
		\hline
		$R_{\text{th}}$ & data rate requirement & $5$ Mbits/s \\
		\hline
		$\omega$ & bandwidth & $1$ MHz \\
		\hline
		$N_{0}$ & noise power spectral density & $4.002\times10^{-18}$ Watts/Hz \\
		\hline
		
	\end{tabular}\label{tabsim}\vspace{-1em}
\end{table}

\begin{figure}[t!]
	\centering
	\centerline{\includegraphics[width=6.5cm]{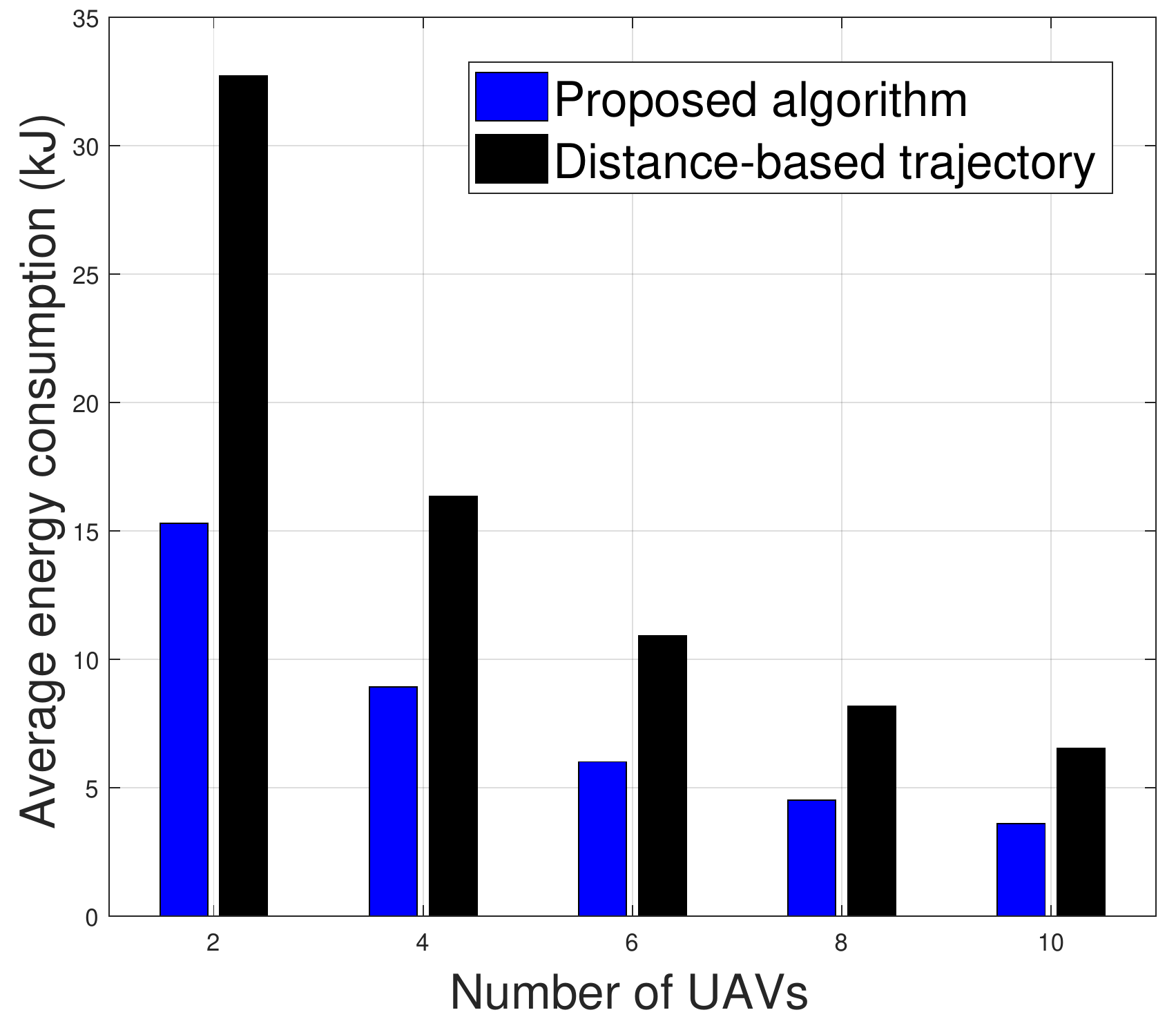}}\vspace{-0.2cm}
	\caption{\small Total energy consumption by the UAV swarm versus the number of UAVs.}\vspace{-.3cm}
	\label{fig1}
\end{figure}

Figure \ref{fig1} compares the total energy consumption of the UAV swarm for the proposed approach with the baseline algorithm, for $N=100$ inspection points. With no energy constraint, the results in Fig.  \ref{fig1} show the merit of exploiting a UAV swarm. That is, as the size of swarm increases, the average total energy consumption decreases. Moreover, it is clear that the proposed algorithm substantially reduces the energy consumptions, by up to $45 \%$ for a swarm of $K=4$ UAVs, compared with the baseline approach. Clearly, the performance gap decreases, as the number of UAVs increases. 

\begin{figure}[t!]
	\centering
	\centerline{\includegraphics[width=6.5cm]{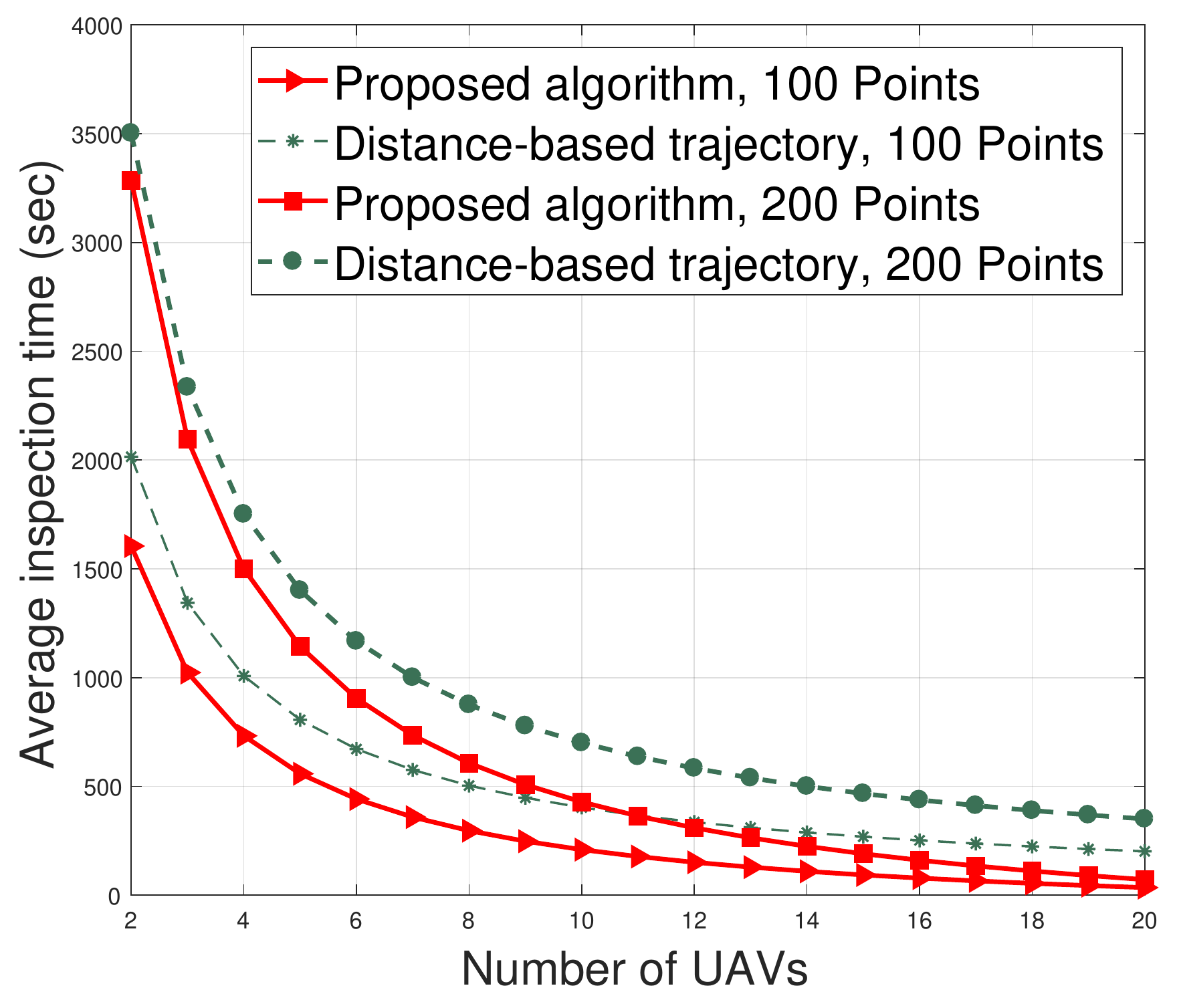}}\vspace{-0.3cm}
	\caption{\small Average inspection time versus the number of UAVs.}\vspace{-.2cm}
	\label{fig2}
\end{figure} 

\begin{figure}[t!]
	\centering
	\centerline{\includegraphics[width=6.5cm]{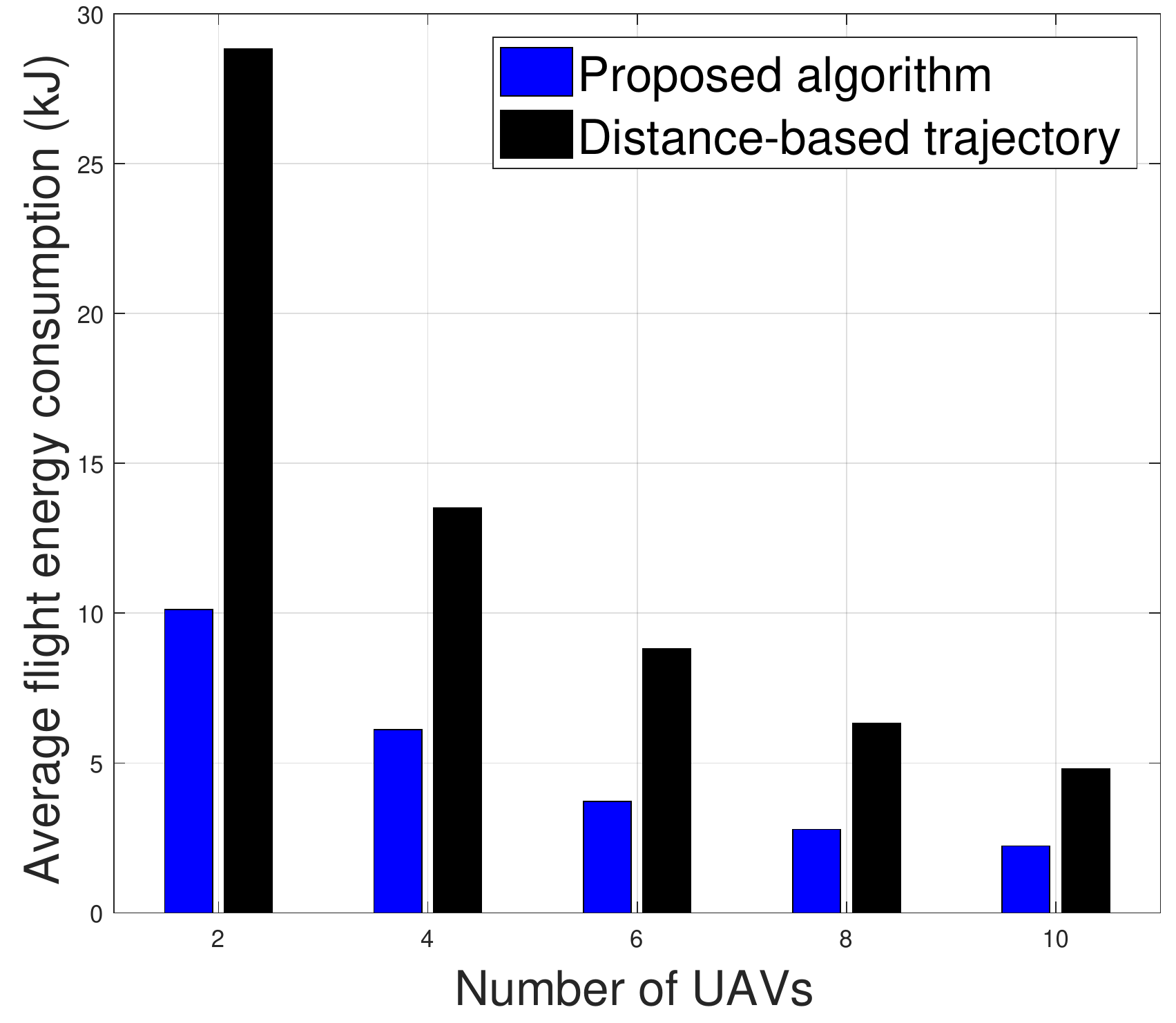}}\vspace{-0.1cm}
	\caption{\small Average flight energy consumption by the UAV swarm versus the number of UAVs.}\vspace{-.5cm}
	\label{fig3}
\end{figure}

Figure \ref{fig2} shows the overall average inspection time for the proposed algorithm, compared with the baseline approach, versus the number of UAVs and for different number of inspection points. Fig. \ref{fig2} shows that as  more UAVs become available to perform inspection, the required time to visit all the inspection points decreases. We note that the rate of decay for the inspection time is faster in the proposed algorithm. Moreover, the proposed approach yields substantial performance gains compared with the distance-based trajectory. For example, for a UAV swarm with $K=12$ UAVs, there are $55\%$ and $47 \%$ reduction in inspection time, respectively, for an area with $N=100$ and $N=200$ inspection points.  This performance gain is due to the fact that the proposed approach accounts for the individual energy constraint of each UAV. As we show in Fig. \ref{fig3}, flying energy constitutes substantial portion of energy consumption. Hence, by minimizing the energy consumption, the proposed approach indirectly minimizes the traversed distance, and thus, the flight time for all UAVs. Such time savings are critical in inspection operations with delay constraints, such as in post-disaster management.

\begin{figure}[t!]
	\centering
	\centerline{\includegraphics[width=6.5cm]{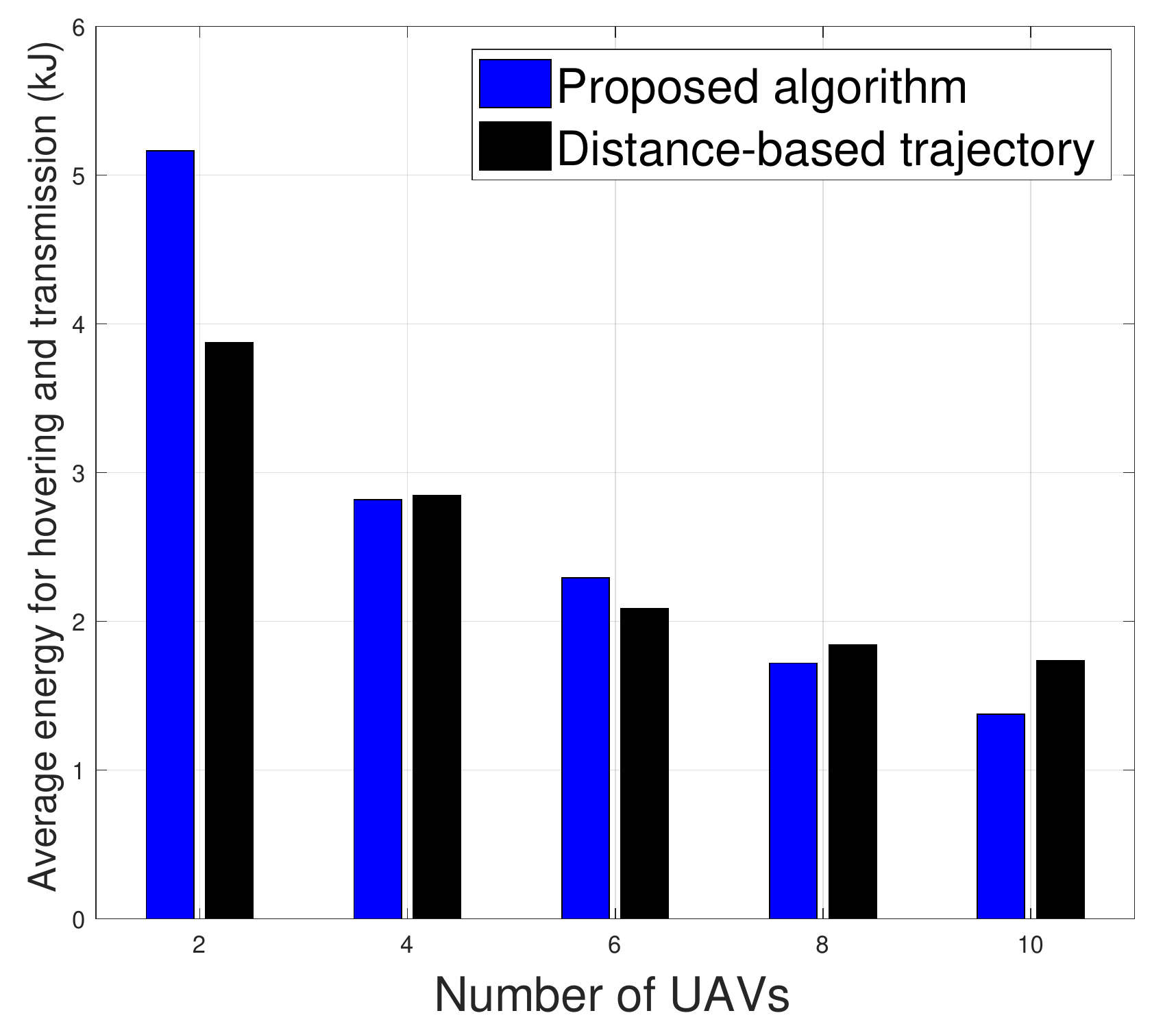}}\vspace{-0.1cm}
	\caption{\small Average hovering plus transmission energy consumption by the UAV swarm versus the number of UAVs.}\vspace{-.4cm}
	\label{fig4}
\end{figure}

Figures \ref{fig3} and \ref{fig4} compare the average energy consumption of the UAV swarm, respectively, for flying and for hovering plus data transmission, for both approaches. From the results in Figs. \ref{fig3} and \ref{fig4},  we can observe that the required energy for flying (i.e., when UAV is mobile) dominates the other energy metrics for hovering and data transmission (i.e., when UAV is static). Comparing the performance of both approaches, we can observe that the flying energy consumption is much higher for the baseline approach. Nonetheless, as shown in Fig. \ref{fig4}, the performance gap is not significant when considering the energy consumption for hovering and data transmission. The main reason is that hovering energy consumption is not highly dependent on the path planning. Moreover, the energy required for data transmissions constitutes small portion of the total energy consumption.

\begin{figure}[t!]
	\centering
	\centerline{\includegraphics[width=6.5cm]{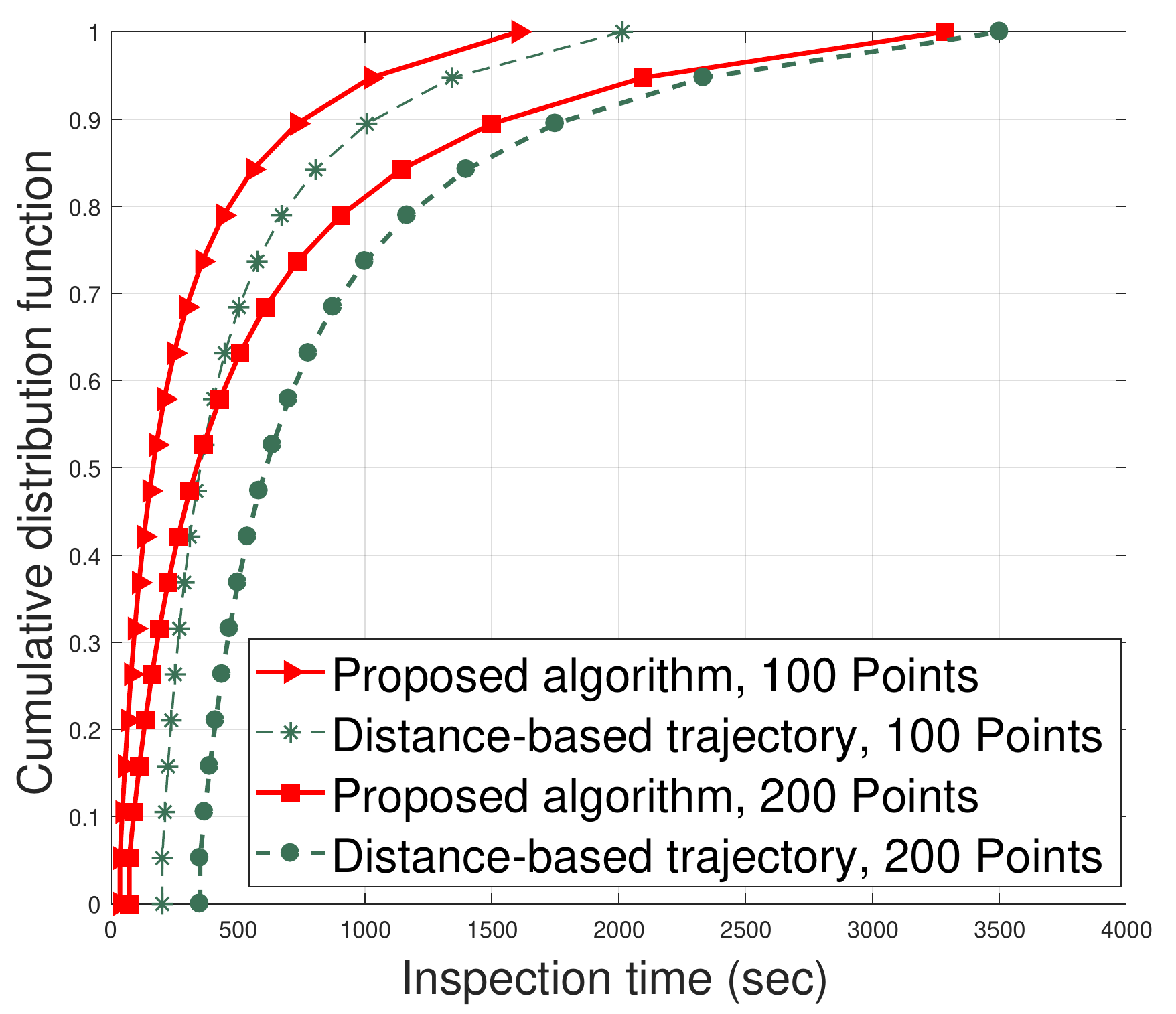}}\vspace{-0.1cm}
	\caption{\small CDF of the inspection time.}\vspace{-.2cm}
	\label{fig5}
\end{figure} 

Figure \ref{fig5} shows the cumulative distribution function (CDF) of the inspection time for $N=100$ and $N=200$ points and with $K=20$ UAVs. This result can capture the reliability of the proposed scheme to perform inspection operation in scenarios with time constraints \cite{magazine}. For example with a target inspection time of $500$ seconds and $N=100$ points, Fig. \ref{fig5} shows that the proposed algorithm can meet this requirement with probability $82 \%$ while there is only  $64 \%$ chance for the baseline approach to satisfy the time constraint. 
Similarly, for $N=200$ and the target inspection time of $900$ seconds, the chance of meeting this requirement is $80\%$ in our approach, whereas $70 \%$ for the distance-based algorithm. 
%Considering both cases, we can certainly come to a point that the reliability in the proposed methodology is higher than the random path allocation approach.

\begin{figure}[t!]
	\centering
	\centerline{\includegraphics[width=6.5cm]{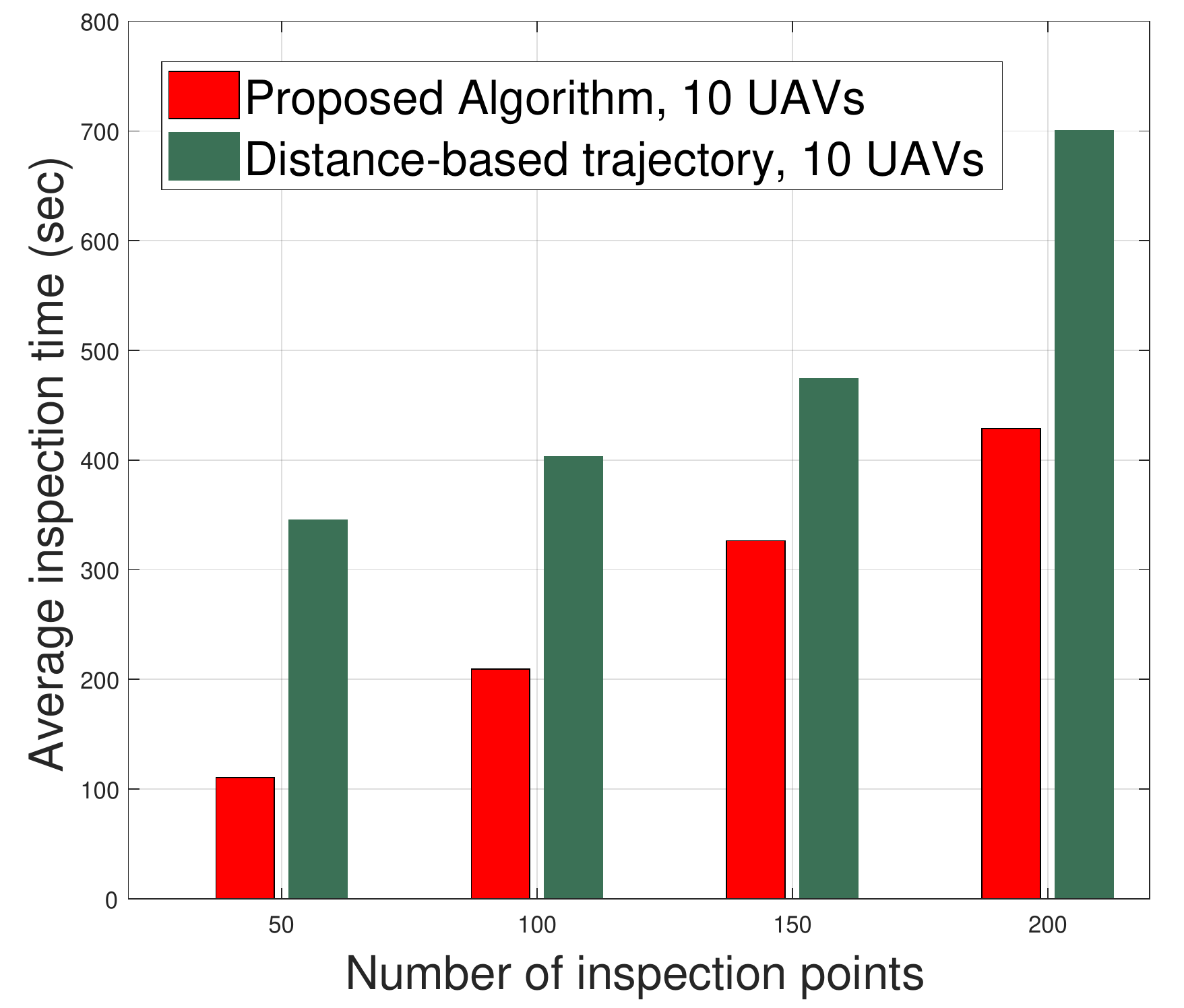}}\vspace{-0.2cm}
	\caption{\small Average inspection time versus number of points.}\vspace{-.2cm}
	\label{fig6}
\end{figure} 

Finally, the average inspection time is shown in Fig. \ref{fig6}, versus the number of inspection points $N$, with $K=10$ UAVs. Fig. \ref{fig6} shows that as there are more inspection points (e.g., in post-disaster scenarios), the average inspection time naturally increases with a fixed number of UAVs. In fact, the results in Fig. \ref{fig6} show that the proposed scheme exploits the UAV swarm more efficiently. For example, for $N=100$, the performance gap between the proposed path planning algorithm and the baseline scheme is $48 \%$.

\section{Conclusions}
In this paper, we have proposed a novel path planning algorithm that minimizes the overall energy consumption by a UAV swarm to autonomously inspect a geographical area. The proposed framework has taken into account different metrics that impact the energy consumption, including flying, hovering, and data transmission by each UAV. We have shown that the proposed algorithm solves the path planning problem in polynomial time, while taking into account the individual energy constraints of the UAVs. Simulation results have shown that the proposed approach substantially outperforms the baseline solution with distance-based path planning, both in terms of required inspection energy and time. 
%\vspace{-.2cm} 
\def\baselinestretch{.94}
\bibliographystyle{IEEEbib}
\bibliography{references,references2}
\end{document}